\documentstyle[12pt]{article}

\def\be{\begin{equation}}
\def\ee{\end{equation}}
\def\bea{\begin{eqnarray}}
\def\eea{\end{eqnarray}}
\def\E{{\bf E}}
\def\B{{\bf B}}
\def\A{{\bf A}}
\def\J{{\bf J}}
\def\k{{\bf k}}
\def\u{{\bf u}}

\begin{document}

\title{Primordial Magnetic Fields that Last?\thanks{Based on 
  a talk at the {\sl 33rd Rencontres de Moriond:
  Fundamental Parameters in Cosmology}, 17-24 January, 1998, Les
  Arcs, France; NSF-ITP/98-072, {\tt astro-ph/9807159}.}}

\author{ Sean M. Carroll$^a$ and George B. Field$^b$ \\
\\
$^a$ Institute for Theoretical Physics, University of
California, \\ Santa Barbara, CA  93106, USA 
\\
$^b$ Harvard-Smithsonian Center for Astrophysics,
60 Garden St., \\ Cambridge, MA  02138, USA }

\date{}

\maketitle

\begin{abstract}
The magnetic fields we observe in galaxies today may have their
origins in the very early universe.  While a number of mechanisms
have been proposed which lead to an appreciable field amplitude
at early times, the subsequent evolution of the field is of crucial
importance, especially whether the correlation length of the field
can grow as large as the size of a protogalaxy.  This talk is a
report on work in progress, in which we consider the fate of one
specific primordial field scenario, driven by pseudoscalar effects
near the electroweak phase transition.  We argue that such a
scenario has a number of attractive features, although it is still
uncertain whether a field of appropriate size can survive until
late times. 
\end{abstract}

\vfill

\eject

\baselineskip=18pt

\section{Introduction}

Observations clearly indicate that galaxies typically possess 
magnetic fields of strength $B\sim 10^{-6}$~Gauss
which are coherent over length scales comparable
to the size of the galaxies \cite{zh}.  At this point it is unclear
whether these fields originate in astrophysical processes operating
during the epoch of galaxy formation and afterward, or can be
traced back to a primordial mechanism in the early universe.
If a primordial mechanism is responsible, it is necessary to
generate a field of amplitude $B \sim 10^{-9}$~G over a comoving 
scale $\lambda \sim 10^6$~pc (the comoving size of a region which
condenses to form a galaxy); such a field can be amplified 
during the process of condensation to the amplitudes observed today.

There are a number of requirements such a scenario must satisfy.
First, it is necessary to generate a field of significant 
size in the early universe, as the field will tend to decay as
$B\propto R^{-2}$ as the universe expands.  (This simple scaling
will be modified when we take into account plasma effects, but the
need for a large initial field will only become more acute.)
Second, the fields must not be significantly
damped in between their formation and the condensation of protogalaxies.
Damping can take different forms, including ordinary Ohmic dissipation
(the exponential decay of electromagnetic fields in plasmas of
finite conductivity) and ``Silk'' damping of MHD modes by photon
and neutrino viscosity.  Third, it is necessary to boost the 
coherence length of the fields, which are typically formed with
much smaller length scales than those of galaxies.  For example,
the comoving horizon size at the electroweak scale (which sets an
upper limit to the correlation length of any field generated 
by a causal mechanism at the electroweak phase transition) is smaller
by a factor of $10^{-10}$ than the comoving length scale associated
with a protogalaxy.  

With these requirements in mind, we will examine the prospects
of a particular scenario for primordial field generation.  Similar
arguments will apply to other possibilities, although we will see
that this scenario has a number of attractive features.

\section{Magnetic Fields from Pseudoscalars}

Consider a pseudoscalar field $\phi$ which couples to electromagnetism
via an interaction Lagrange density
\be
  {\cal L} = \phi F_{\mu\nu} \widetilde F^{\mu\nu}\ ,
\ee
where the dual field strength tensor is defined by
$\widetilde F^{\mu\nu} = {1\over 2} \epsilon^{\mu\nu\rho\sigma}
F_{\rho\sigma}$.  Such an interaction leads to a modified form of 
Maxwell's equations, given by
\bea
  -\partial_t \E + \nabla\times\B & = & 4\pi\J - \dot\phi \B
  + \nabla\phi \times\E \\
  \nabla\cdot\E & = & 4\pi\rho - \nabla\phi \cdot\B \\
  \partial_t \B + \nabla\times\E & = & 0 \\
  \nabla\cdot\B & = & 0\ .
\eea
(For simplicity we are working in flat spacetime, but the
generalization to Robertson-Walker universes is straightforward.)
In a conducting plasma we will also have Ohm's Law, $\J = \sigma\E$,
where $\sigma$ is the conductivity.

We assume that the pseudoscalar is spatially homogeneous, so $\nabla\phi$
can be neglected.  We also drop $\rho$, assuming there is no net
charge density, and $\partial_t\E$, as time variations in
the electric field will be small.  Under these assumptions we derive
an equation for the magnetic field in Fourier space,
\be
  \partial_t \B = -{1\over{4\pi\sigma}}(k^2 \B + i\dot\phi \k\times\B)
  \ .
\ee
This equation can be analyzed by decomposing $\B$ into orthonormal modes
perpendicular to the wavevector $k$, $\B(\k) = b_1\u_1 + b_2\u_2$.  In
fact, it is most convenient to work with circularly polarized modes,
$b_\pm = b_1 \pm ib_2$.  We can then solve explicitly for the time
evolution:
\be
  b_\pm(t,\k) = b_\pm(0,\k) \exp\left[-{k\over{4\pi\sigma}}
  \left(kt\pm \int \dot\phi\, dt\right)\right]\ .
\ee
We see that the $b_-$ modes can grow exponentially if $\dot\phi > k$,
with maximum growth for $k={1\over 2}\dot\phi$.

The existence of such an exponentially growing mode suggests that
such a pseudoscalar could generate large field strengths in the early
universe.  It remains to specify an identity and dynamics for the
$\phi$ field itself.  A scenario along these lines was proposed by
Turner and Widrow \cite{tw}, who noted that the axion was a particle
with appropriate couplings, but did not analyze the possibility in
detail.  Garretson, Field and Carroll \cite{gfc} considered a generic 
pseudo-Goldstone boson evolving during inflation, and found that it was
not possible to generate fields which were both of sufficient strength
and interestingly large length scales.  More recently, Joyce and
Shaposhnikov~\cite{js} note that a chemical potential for right-handed 
electron number, generated by processes at the grand unification scale,
interacts with electromagnetism in an equivalent fashion, if we
simply identify the chemical potential with $\dot\phi$.  (See also
the work of Cornwall \cite{cornwall} and Son \cite{son}.)

The Joyce and Shaposhnikov scenario, which involves only standard
electroweak physics once the chemical potential is generated, is
less flexible than a generic pseudoscalar boson and accordingly more
predictive.  For definiteness we will consider the fate of the magnetic
fields generated by such a mechanism, although other pseudoscalars
would have very similar effects.  Joyce and Shaposhnikov estimate
that their scenario can lead to magnetic fields of order $B\sim 10^{22}$~G
on a length scale $\lambda \sim 10^{-8} H^{-1}_{EW}$, where $H_{EW}$
is the Hubble parameter at the electroweak scale.  As this is 18 
orders of magnitude smaller than the desired comoving length scale,
we must seek a mechanism for increasing the coherence length of the
field.

\section{The Inverse Cascade}

A well-known property of ordinary hydrodynamical turbulence is
the cascade of energy, injected at a certain length scale, down
to smaller scales.  In MHD, however, magnetic energy can both
cascade to small scales and inverse cascade to large scales.
This phenomenon was investigated numerically and analytically in
the 1970's and 1980's by Pouquet and collaborators \cite{pouquet}, and
has been advocated as an important factor in the evolution of
primordial magnetic fields by Brandenburg, Enqvist and Olesen
\cite{beo}.

In order for an inverse cascade to be operative, two requirements
must be met:  an injection of turbulence into the medium, and
a nonvanishing expectation value for
some pseudoscalar quantity.  Turbulence, which is needed to transfer
energy between disparate length scales, can arise from (for example)
bubble collisions during a first-order electroweak phase transition.
We will assume that this is the case, although little is known about
the order of the electroweak transition in the real world.  The
necessity of a nonvanishing pseudoscalar can be seen by considering the
time variation of the magnetic field, which satisfies 
$\partial_t\B(\k) \sim \k\times\E$.  To obtain exponential growth in
$\B$ at a wavenumber $k$, we require $\E(\k) = \alpha(\k) \B(\k)$,
where $\alpha$ is a manifestly pseudoscalar coefficient.
Numerical simulations by Pouquet et al. \cite{pouquet} have verified
the existence of an inverse cascade if and only if the configuration
possesses significant magnetic helicity, $H^M = \int \A \cdot \B \, d^3x$.

The pseudoscalar
mechanism discussed in the previous section creates a field with
maximal magnetic helicity.  The circularly
polarized modes $b_\pm$ are of opposite helicity, and one will be
suppressed while the other is amplified as $\phi$ evolves; the
resulting field is maximally helical.  Hence, such a field should
have magnetic energy transferred to larger scales, while we would
not expect such behavior in a generic situation.

The transfer of energy to large scales is typically very efficient;
numerical simulations indicate~\cite{pouquet} that the scale increases 
linearly by a factor $\Delta t/t_{\rm turb}$, where the turbulence 
timescale $t_{\rm turb}$ can be taken (somewhat optimistically) to be 
$T_{\rm EW}^{-1} \sim 10^{-15} H_{\rm EW}^{-1}$.  Since the naive
picture of linear growth would increase the coherence length beyond
the Hubble distance, we can assume that the growth would saturate at
that point.  In the (once again optimistic, but not implausible) 
assumption that the field strength is undiminished by this process, 
we are therefore left with a field of amplitude $B\sim 10^{22}$~G and
a length scale $\lambda\sim H_{\rm EW}^{-1}$ (a comoving scale
of $10^{-4}$~pc).

\section{Subsequent Evolution}

A potentially important role in the evolution of the field from the
electroweak scale to today can be played by the damping of MHD modes
by photon and neutrino viscosity \cite{jko}.  This damping can
dramatically decrease the amplitude of primordial magnetic fields
before they have a chance to form galactic fields.  However,
these modes do not necessarily represent oscillations around a
zero-field configuration, but around a force-free field, for which
$\J\times\B =0$.  For the situation under consideration, such
damping will not be important, as numerical simulations \cite{pouquet}
(as well as experiments in tokamaks \cite{os}) indicate that the result 
of an inverse cascade is a force-free field configuration.

This leaves us with the question of whether the length scale of
the magnetic field can be expanded to the dimensions of a protogalaxy.
Presumably an inverse cascade mechanism cannot be very helpful,
as there is likely to be no source of turbulence subsequent to the
electroweak phase transition.  A pessimistic scenario would imagine
that the field is frozen in and expands with the universe, redshifting
as $R^{-2}$.  We would then be left with a field amplitude
$B\sim 10^{-8}$~G on a scale of $10^{-4}$~pc today.  To estimate the
amplitude on galactic scales, we can consider the incoherent
superposition of fields in uncorrelated domains.  Since the resulting 
field goes like one over the square root of the number of domains,
which in turn goes as the volume of the region under consideration,
the field on megaparsec scales is 
\be
  B({\rm Mpc}) \sim \left({{1~{\rm Mpc}}\over{10^{-4}~{\rm pc}}}
  \right)^{-3/2}B(10^{-4}~{\rm pc}) \sim 10^{-23}~{\rm G}\ .
\ee
This is much less than the sought-after amplitude $10^{-9}$~G, so
it is necessary to imagine a more active mechanism for increasing 
the coherence length.

In fact, the coherence length will certainly grow faster, as
uncorrelated domains come into causal contact and magnetic field
lines smooth themselves out \cite{hdd}.  At this point we do not
have a reliable estimate of the rate at which this happens, nor
of the potential dilution of the field strength during this process.
Instead, we can proceed under optimistic assumptions to see whether
there is any prospect of generating the required field.

The optimistic expectation we consider is that the field
rearranges itself at the Alfv\'en speed,
\be
  v_A = \left({{\rho_B}\over{\rho_{\rm tot}}}\right)^{1/2}\ ,
\ee
which has the value $v_A\sim 10^{-2}$ during the radiation
dominated era (for the parameters used above) and $10^{-2}T$(eV)
during matter domination.  The correlation length will then obey
\be
  {{d\lambda}\over {dt}} = H\lambda + v_A\ ,
\ee
representing the separate effects of Hubble expansion and 
Alfv\'en rearrangement.  Plugging in the appropriate numbers,
we find that the Alfv\'en speed dominates until the time of
matter-radiation equality, after which the Hubble expansion is
most important (so that the comoving length remains constant).
It so happens that the Hubble size at matter-radiation equality,
which will characterize the correlation length of the magnetic
field under these assumptions, corresponds to a comoving scale
of 1~Mpc, nicely consistent with the size of a protogalaxy.

Under our optimistic scenario, then, we are able to generate
fields of the appropriate length scale, but no larger.  Under
the additional optimistic assumption that the amplitude of the
field remains undiminished as it smooths out, we obtain
$B(1~{\rm Mpc})\sim 10^{-8}$~G, just as required for the 
primordial scenario.

The fact that we are only able to achieve this result under
such optimistic conditions is somewhat discouraging.  A more
realistic calculation would include both the fact that viscosity
will act to retard the Alfv\'en rearrangement, and that 
dissipation will act to diminish the amplitude of the field.
At this point we are not confident in our understanding of the
magnitude of these effects; work in this direction is in progress.
While the prospects for primordial mechanisms for magnetic
field generation do not seem hopeful in light of this analysis,
there is still a chance that a more careful examination will
reveal that our optimistic assumptions are actually warranted.
The importance of this topic justifies a concerted effort to
understand whether this possibility can be realized.

\section*{Acknowledgments}
This work was supported in part by the National Science Foundation
under grant PHY/94-07195 and NASA under grant NAGW-931.

%\section*{References}


\begin{thebibliography}{99}

\bibitem{zh} E.G. Zweibel and C. Heiles, {\it Nature} {\bf 385}, 131 (1997).

\bibitem{tw} M.S. Turner and L.M. Widrow, {\it Phys. Rev. D} {\bf 37},
  2743 (1988).

\bibitem{gfc} W.D. Garretson, G.B. Field, and S.M. Carroll,
  {\it Phys. Rev. D} {\bf 46}, 5346 (1992); {\tt hep-ph/9209238}.

\bibitem{js} M. Joyce and M. Shaposhnikov, {\it Phys. Rev. Lett.} {\bf 79}, 
  1193 (1997); {\tt astro-ph/9703005}.

\bibitem{cornwall} J.M. Cornwall, {\it Phys. Rev. D} {\bf 56}, 6146 (1997);
  {\tt hep-th/9704022}.

\bibitem{son} D.T. Son, preprint MIT-CTP-2724 (1998); {\tt hep-ph/9803412}.

\bibitem{pouquet} U. Frisch, A. Pouquet, J. L\'eorat, and A. Mazure,
  {\it J. Fluid Mech.} {\bf 68}, 769 (1975);  A. Pouquet, U. Frisch, 
  and  J. L\'eorat, {\it J. Fluid Mech.} {\bf 77}, 321 (1976);
  A. Pouquet and G.S. Patterson, {\it J. Fluid Mech.} {\bf 85}, 305
  (1978);  M. Meneguzzi, U. Frisch and A. Pouquet, {\it Phys. Rev.
  Lett.} {\bf 47}, 1060 (1981).

\bibitem{beo} A. Brandenburg, K. Enqvist and P. Olesen, {\it Phys. Rev.
  D} {\bf 54}, 1291 (1996), {\tt hep-ph/9608422};  P. Olesen,
  {\tt astro-ph/9610154};  K. Enqvist, {\tt astro-ph/9707300}.

\bibitem{jko} K. Jedamzik, V. Katalini\'c, and A.V. Olinto,
  {\it Phys. Rev. D} {\bf 57}, 3264 (1998); 
  {\tt astro-ph/9606080}.

\bibitem{os} S. Ortolani and D.D. Schnack, {\it Magnetohydrodynamics of
  Plasma Relaxation}, World Scientific (1993).

\bibitem{hdd}C. Hogan, {\it Phys. Rev. Lett.} {\bf 51}, 1488 (1983);
   K. Dimopoulos and A.-C. Davis, {\it Phys. Lett.} {\bf B390}, 87 (1997),
  {\tt astro-ph/9610013}.



\end{thebibliography}
\end{document}